\documentclass[11pt]{article}

\usepackage[margin=2.15cm]{geometry}
\usepackage{graphicx}
\usepackage{setspace}
\usepackage{titlesec}
\usepackage{fancyhdr}
\usepackage{url}
\usepackage[hidelinks]{hyperref}
\usepackage{caption}
\usepackage{xcolor}

\usepackage{multicol}

\usepackage[backend=biber,
            style=numeric,
            maxnames=2,
            minnames=1,
            doi=false,
            sorting=none,
            url=false,
            isbn=false,
            giveninits=true,
            eprint=false]{biblatex}
\DeclareBibliographyDriver{article}{%
  \usebibmacro{author/translator+others}%
  \setunit{\addspace}%
  \printfield{year}%
  \setunit{\addcomma\space}%
  \printfield{journaltitle}%
  \setunit{\addspace}%
  \printfield{volume}%
  \setunit{\addcomma\space}%
  \printfield{pages}%
  \finentry
}

\AtEveryBibitem{%
  \clearfield{month}%
  \clearfield{day}%
  \clearfield{title}%
  \clearfield{subtitle}%
  \clearfield{note}%
  \clearfield{pages}%
  \clearfield{editor}%
  \clearlist{language}%
}

\titleformat{\section}{\large\bfseries}{\thesection}{1em}{}
\titleformat{\subsection}{\normalsize\bfseries}{\thesubsection}{1em}{}
\titlespacing*{\section}{0pt}{0.5em}{0.0em}%{1.2ex plus 0.5ex minus 0.2ex}{0.8ex}
  
\titlespacing*{\subsection}{0pt}{1.0ex plus 0.3ex minus 0.2ex}{0.6ex}

\title{\Huge \textbf{Connecting galaxies with their haloes - from parsec to Mpc scales}}
\author{
\large K. Fahrion\textsuperscript{1}, J. van de Sande\textsuperscript{2}, K. R. Akhil\textsuperscript{3}, M. A. Beasley\textsuperscript{4, 5, 6}, F. Belfiore\textsuperscript{7}, \\ M. L. L. Dantas\textsuperscript{8},  P. K. Das\textsuperscript{9}, E. Emsellem\textsuperscript{7}, J. Hartke\textsuperscript{10}, M. Hilker\textsuperscript{7}, A. Monreal-Ibero\textsuperscript{11}, \\ A. Prieto\textsuperscript{4, 5}, M. Raouf\textsuperscript{11, 12}, S. Thater\textsuperscript{1}, I. Zinchenko\textsuperscript{13}\\
\small \textsuperscript{1}University of Vienna, Austria, 
\small \textsuperscript{2}University of New South Wales, Australia,\\
\small \textsuperscript{3}Indian Institute of Astrophysics, Bangalore, India,
\small \textsuperscript{4}Instituto de Astrofísica de Canarias, Spain, \\
\small \textsuperscript{5}Universidad de La Laguna, Spain, 
\small \textsuperscript{6}Centre for Astrophysics and Supercomputing, \\\small Swinburne University, Australia, 
\small \textsuperscript{7}European Southern Observatory, Germany, \\
\small \textsuperscript{8}Pontificia Universidad Católica de Chile, 
\small \textsuperscript{9}University of Queensland, Brisbane, Australia, \\
\small \textsuperscript{10}University of Turku, Finland, 
\small \textsuperscript{11}Leiden Observatory, Netherlands \\
\small \textsuperscript{12}Delft University, Netherlands 
\small \textsuperscript{13}Heidelberg University, Germany
\\
\\
\small Contact: K. Fahrion \& J. van de Sande
}

\date{\small \today}

\begin{document}

% --- Cover Page ---
\maketitle
\thispagestyle{empty}
\vfill
\begin{center}
    \textbf{Abstract} \\[0.5em]
    \begin{minipage}{0.90\textwidth}
    Galaxy evolution is driven by processes occurring across a wide range of scales, from star formation within giant molecular clouds (parsec scales) to outflows and secular evolution across entire galaxies (kpc scales), and the interplay between galaxies, their dark matter haloes, and large-scale structures (Mpc scales). Connecting the distribution of baryonic matter and energy across these scales will remain one of the key challenges for both theoretical and observational astrophysics in the coming decade.
    A major development towards meeting this challenge has been the growing ability to obtain highly spatially resolved (parsec-scale) integral-field spectroscopic observations (e.g. with VLT/MUSE), as well as to probe the extremely low-surface brightness outskirts of galaxies at large radii and high vertical scale heights. To combine the two regimes, we need a paradigm shift in the way we do spectroscopy on galaxies, especially considering the ongoing and future photometric surveys.
   The next decade will also bring a revolution in extensive photometric surveys of large areas of the sky, uncovering low surface brightness features around nearby galaxies. However, to fully understand the processes that connect galaxies to their haloes, shape low surface brightness features, and drive secular evolution, spatially resolved spectroscopy will be essential. Here, we outline the need for wide-field spectroscopic observations of statistically significant samples of nearby galaxies and highlight the key questions that can only be addressed with such data.    
    \end{minipage}
\end{center}
\vfill
\newpage

% --- Page 1 ---
\section{Context}
A full understanding of how galaxies of all morphological types assemble, form stars, and interact with their large-scale environments requires a fundamental shift in observational strategy. Key questions of how gas flows shape galaxy growth, how environment influences internal structure, and how star formation proceeds from cloud scales to halos cannot be answered with single wavelength or limited field-of-view data. What is needed is a new paradigm: wide-field, multi-wavelength imaging combined with equally wide-field, spatially resolved spectroscopy across large, representative samples of nearby galaxies ($<$ 50 Mpc).
Upcoming facilities provide the basis for this transformation. Current observatories (VLT, ALMA, JWST, Euclid, \cite{Leroy2021, Williams2024}) and upcoming ones such as Rubin LSST and the Roman Space Telescope, along with new UV missions (UVEX \cite{Kulkarni2021}; CASTOR \cite{Cote2019}; LUVOIR \cite{Tumlinson2019}), next-generation X-ray telescopes (e.g. NewAthena \cite{NewAthena}), and SKA 21-cm mapping, will deliver high-resolution, multi-wavelength views of galaxies from their disks to their faint outskirts. These surveys will offer the statistical sample sizes covering hundreds of galaxies and spatial resolution needed to probe substructure assembly and cloud-scale star formation ($\sim$ 10–100 pc).
However, only when paired with wide-field, spatially resolved (integral-field spectroscopy, IFS) and multi-object spectroscopy (MOS) at optical wavelengths can these datasets reveal the full baryon cycle and its dependence on galaxy assembly and environment. The following sections outline the science questions that demand this integrated approach.

\section{Open questions}
\textbf{How far is the reach of galactic fountains and how do they regulate star formation on a parsec scale?} 
This question hinges on measuring the interplay between stellar feedback, chemical enrichment, and cloud-scale star formation across entire galactic discs \cite{ErrozFerrer2019, McLeod2021}. At scales below $\sim$100 pc, where star-forming regions, HII regions and star clusters are individually resolved, feedback processes determine whether gas is expelled, recycled, or collapses, and classical integrated scaling relations no longer apply \cite{KennicuttEvans2012, Chevance2022}. External accretion from the circumgalactic medium may also supply gas to galaxies, representing a distinct process from internally driven fountain cycles. Deep imaging and HI data reveal narrow dust and gas filaments that appear to funnel material from tens of kiloparsecs into the disc and even the central regions \cite{Prieto2019}. To establish the spatial extent of these processes, we must map ionisation conditions, feedback pressures, metal abundances, and escape fractions across thousands of regions spanning the full dynamical range of galactic environments.
Crucially, these diagnostics must extend to galaxy outskirts \cite{Ferguson1998}, where existing constraints on chemical abundances and gas-phase physics remain sparse and limited to individual systems and regions. Deep, wide-field integral field spectroscopy is required to obtain gas metallicities and to trace how enriched gas is mixed or transported into low-density outer discs \cite{KadoFong2020}. Only a survey providing spectroscopy for thousands of regions in hundreds of galaxies can provide insights into chemical enrichment, diffusion, and the role of morphological features such as spiral arms in chemical mixing.

\textbf{To what degree do the building blocks of the halo differ from, or align with, those of other galaxy components?} 
Answering this question requires disentangling the chemo-dynamical signatures of components that originate through fundamentally different pathways. While galaxy discs and central regions might form mainly through internal processes, halo regions likely comprise both heated disc stars and accreted populations. Stellar population and orbit modelling now allow for the detailed reconstruction of the origin of individual components in a few galaxies (e.g. \cite{Gadotti2019, Poci2021, Emsellem2022}). Still, current spectroscopic samples (e.g. CALIFA \cite{CALIFA}, MaNGA \cite{MaNGA}) lack the uniform spatial coverage and statistical size needed to compare these components across the full galaxy population. 
A volume-limited, wide-field spectroscopic survey is essential to determine whether halo stars, GCs, tidal debris, outer discs, and inner structural components share similar enrichment histories or arise from different assembly mechanisms. Deep IFS of galaxy outskirts will capture the poorly understood physics of low-density gas, tidal features, and extended discs, while resolving age- and metallicity gradients across all components. Spatially resolved angular-momentum studies further show that connecting disc structure to halo assembly requires velocity fields and emission-line diagnostics across entire galaxies, including their faint outskirts \cite{Dutton2012}, and the dynamical relaxation state of haloes characterised through indicators such as luminosity gaps, centroid offsets, or kinematic misalignments also provides a complementary probe of their formation time and assembly history \cite{Raouf2014}. Incorporating tracers of dynamical ages into wide-field spectroscopic surveys will connect halo assembly, disc growth, and the evolutionary state of central galaxies. Only such comprehensive spectroscopy can distinguish in situ from accreted structures, quantify the role of environment, and reveal how the building blocks of haloes relate to those shaping discs and bulges.

\textbf{How does large-scale filamentary structure shape the outer haloes of galaxies? Where do mergers and streams deposit their stars and gas?} 
Upcoming imaging surveys will unveil a wealth of faint substructures, such as streams, shells and tidal tails that trace a galaxy’s recent and ancient interactions with its environment \cite{Martin2022, Urbano2025, Romanowsky2025}. Yet the key questions remain unanswered: Where do stripped satellites deposit their stars? Do haloes build up smoothly or through discrete, chemically distinct events? How does the large-scale structure affect galaxy assembly?

To address these questions beyond the Milky Way, spectroscopy is required. Wide-field IFS enables measurements of ionised gas in outer discs and tidal debris, thereby providing insights into their origin \cite{Fensch2020}. At the same time, multi-object spectroscopy observations of satellites, globular clusters, and planetary nebulae provide complementary tracers of the mass distribution and chemo-dynamical assembly of haloes (e.g. SLUGGS \cite{Brodie2014}, ePN.s \cite{Pulsoni2018}). Together with absorption-line spectroscopy toward background quasars, these observations trace inflows, outflows, and enrichment of the circumgalactic medium. This combination directly tests $\Lambda$CDM predictions for accretion, satellite planes, and the hierarchical growth of haloes \cite{Pawlowski2024}, providing a comprehensive picture of how large-scale filamentary structure shapes the outer regions of galaxies and governs the deposition of merger debris.

\section{Technical requirements and synergies}
A facility capable of delivering deep, spatially resolved spectroscopy across entire nearby galaxies ($<$ 50 Mpc) and their outskirts must combine large collecting area (mirror size 12 - 15m), wide-field IFS capabilities, and simultaneous multi-object spectroscopy. The scientific goals outlined above, which involve connecting galaxy components with their haloes in a statistically significant sample, cannot be met with existing instruments. MUSE and BlueMUSE provide exquisite spatial resolution, but their fields of view limit survey speed and prohibit uniform coverage of full discs and haloes in large samples. Likewise, their sensitivity is insufficient to reach the extremely low surface-brightness regimes where outer-disc ionised gas, tidal debris, and faint halo populations reside. No existing instrument offers wide-field IFS with MOS and high multiplexing, making a next-generation facility crucial for connecting galaxies with their haloes.

A wide-field IFS/MOS facility would provide essential spectroscopic context to the major observatories operating in the coming decades. For example, SKA will map the distribution and kinematics of neutral hydrogen to unprecedented radii \cite{SKA}; combining these data with optical spectroscopy will reveal how the neutral and ionised phases connect and how galaxies regulate their gaseous boundaries. NewAthena and other future X-ray missions will characterise the hot circumgalactic medium \cite{NewAthena}, and spectroscopic mapping of ionised gas and halo tracers will link this hot reservoir to the warm and cold phases that participate in the baryon cycle. Synergies with the ELT will be particularly powerful: 
Instruments such as HARMONI or MOSAIC will deliver unparalleled depth and spatial resolution in a few selected regions, and will benefit directly from wide-field measurements that constrain stellar masses, large-scale gas flows, and the dynamical structure of entire galaxies (e.g. \cite{Thater2019}). Finally, deep imaging from Rubin LSST, Euclid, the Roman Space Telescope, and ARRAKHIS will uncover low-surface brightness structures such as faint outskirts and tidal debris around nearby galaxies \cite{Martin2022, Urbano2025, Arrakhis, Romanowsky2025}, and complementary wide-field spectroscopy will be essential for enabling robust reconstruction of galaxies’ merger histories. Together, these synergies will enable a complete, multi-phase view of how galaxies form, evolve, and interact with their surroundings.

\section{Conclusions}
Full and detailed understanding of the physical processes that shape galaxies and connect them to their dark matter halo will require a wide-field spectroscopic facility to complement the rich imaging surveys of the next decades. By enabling deep, spatially resolved spectroscopy from galaxy centres to their faintest outskirts and haloes, it would open a new parameter space for understanding the baryon cycle, hierarchical assembly, and environmental influence. Such a facility would place ESO at the forefront of galaxy evolution studies for decades to come.

\begin{multicols}{2}
\printbibliography
\end{multicols}

\end{document}